# Antecedents of Consumer Regret Frequency: The Roles of Decision Agency, Status Signaling, and Online Shopping Preference

**Shawn Berry, DBA[1]***

**December 30, 2025**

[1]William Howard Taft University, Lakewood CO, USA

*Correspondence: shawnpberry@gmail.com

## Abstract

Consumer regret is a widespread post-purchase emotion that significantly impacts satisfaction, product returns, complaint behavior, and customer loyalty. Despite its prevalence, there is a limited understanding of why certain consumers experience regret more frequently as a chronic aspect of their engagement in the marketplace. This study explores the antecedents of consumer regret frequency by integrating decision agency, status signaling motivations, and online shopping preferences into a cohesive framework. By analyzing survey data (n=338), we assess whether consumers' perceived agency and decision-making orientation correlate with the frequency of regret, and whether tendencies towards status-related consumption and preferences for online shopping environments exacerbate regret through mechanisms such as increased social comparison, expanded choice sets, and continuous exposure to alternative offers. The findings reveal that regret frequency is significantly linked to individual differences in decision-related orientations and status signaling, with a preference for online shopping further contributing to regret-prone consumption behaviors. These results extend the scope of regret and cognitive dissonance research beyond isolated decision episodes by emphasizing regret frequency as a persistent consumer outcome. From a managerial standpoint, the findings suggest that retailers can alleviate regret-driven dissatisfaction by enhancing decision support, minimizing choice overload, and developing post-purchase reassurance strategies tailored to segments prone to regret..

## 1. Introduction

### 1.1 Background and Rationale

Consumer regret constitutes a widespread emotional response that emerges when individuals retrospectively assess a decision unfavorably in comparison to alternatives that are not chosen, often accompanied by self-reproach and counterfactual contemplation regarding potential alternative choices (Inman & Zeelenberg, 2002; Zeelenberg & Pieters, 2007). Within consumer markets, this emotion is closely aligned with what is commonly referred to as a buyer's remorse, typically occurring post-purchase when the actual performance, value, or symbolic significance of a product does not meet expectations or when superior alternatives become apparent after the fact (Sweeney, Hausknecht, & Soutar, 2000; Tsiros & Mittal, 2000). Regret is also conceptually linked to cognitive dissonance, the psychological discomfort arising from holding conflicting cognitions, such as selecting an option later perceived as inferior, which consumers may attempt to alleviate by modifying attitudes, seeking confirmatory information, or returning products (Sweeney et al., 2000). Although these terms are occasionally used interchangeably, regret specifically emphasizes the counterfactual comparison between chosen and forgone options, whereas cognitive dissonance broadly encompasses tension among beliefs and behaviors. From a managerial standpoint, frequent consumer regret poses challenges, as it diminishes post-purchase satisfaction, increases product returns, and may incite complaint behavior, negative word-of-mouth, and brand switching (Chebat, Davidow, & Codjovi, 2005; Tsiros & Mittal, 2000; Zeelenberg & Pieters, 2004). Retailers and service providers invest significantly in acquisition and promotion activities. However, the downstream effects of regret can undermine the long-term value of these investments by eroding trust and loyalty, particularly in competitive categories in which dissatisfied consumers can readily switch to alternatives. Regret also complicates the consumer decision-making process itself; anticipation of potential future regret may lead consumers to over-deliberate, postpone decisions, or seek excessive reassurance, thereby increasing perceived decision difficulty and potentially diminishing choice confidence (Inman & Zeelenberg, 2002; Zeelenberg & Pieters, 2007). In digital environments with abundant information and alternatives, this anticipation and experience of regret may intensify through continuous exposure to better deals, new products, and social comparison cues.

Despite extensive research on the antecedents and consequences of regret in specific decision-making contexts, significant gaps persist in our understanding of the frequency with which consumers experience regret as a chronic aspect of marketplace interaction. Much of the existing literature concentrates on isolated decision episodes, experimental scenarios, or narrowly defined post-purchase contexts, typically assessing regret at a single point in time and in relation to a specific choice (e.g., Tsiros & Mittal, 2000; Inman & Zeelenberg, 2002). Considerably less attention has been devoted to the frequency with which consumers report regret across everyday purchases and to the broader array of psychological traits, decision orientations, and shopping practices that may systematically influence this chronic regret frequency. Notably, there is limited knowledge regarding how perceived decision and financial agency, status signaling motives, and online commerce readiness collectively contribute to the frequency with which consumers "definitely" regret purchases, independent of more traditional demographic and deal-seeking variables.

This study addresses these gaps by developing and testing a structural equation model of consumer regret frequency (REGRETFREQ) operationalized on a scale from 1 (definitely no regret) to 5 (definitely always regret). First, we specify and validate a multidimensional measurement model for three key latent constructs–status-signaling orientation, decision/financial agency, and online commerce readiness–assessing their internal consistency and convergent validity using established psychometric criteria. Second, we estimate a structural model that links these latent constructs, along with a comprehensive set of behavioral, attitudinal, and demographic indicators, to the frequency of consumer regret, thereby identifying which factors uniquely predict chronic regret once other influences are controlled. Third, we evaluate the global fit of the proposed model and derive theoretical and managerial implications for how marketers and policymakers might design decision environments, communication strategies, and digital interfaces that enhance consumer agency and quality-focused decision-making and, in turn, reduce the incidence of buyers' remorse and regret-related dissatisfaction. Collectively, these steps aim to provide a more comprehensive and process-oriented account of consumer regret frequency than is currently available in the literature.

## 1.2 Literature Review

Exploring the emotional dynamics that occur after a purchase is a vital aspect of consumer behavior research, with consumer regret standing out as a key concept that presents considerable difficulties for retailers and affects the consumer decision-making process (Sarwar, Awang, & Habib, 2019). Consumer regret is formally described as a negative emotion rooted in cognition that emerges when a consumer realizes or envisions that their current situation could have been improved if they had opted for a different choice (Tsiros & Mittal, 2000). This definition underscores the fundamental mechanism of regret: a counterfactual comparison between the selected option and the option that was not chosen (Zeelenberg and Pieters, 2007). Although often used interchangeably in everyday conversation, academic literature draws clear distinctions between regret and similar post-purchase experiences, such as buyer's remorse and cognitive dissonance. Buyer's remorse is a widely recognized term that typically describes the anxiety or regret felt after making a purchase, especially a costly one (Knezevic, 2024). In marketing, it is frequently equated with post-purchase regret (AlSharhan, 2017). In contrast, cognitive dissonance, a concept introduced by Festinger (2017), represents a unique psychological condition. It refers to the mental discomfort that arises from holding two contradictory beliefs (e.g., "I am intelligent" and "I just made a foolish purchase") (Costanzo, 2013). While regret is an emotion based on comparing what happened with what could have been, dissonance is a motivational state that compels consumers to resolve a conflict, often by rationalizing or seeking information that supports their decision (Zhang, 2018). Regret and dissonance are closely connected, as experiencing regret can lead to dissonance, and resolving dissonance can alleviate regret ( Awwad, Ibrahim, & George, 2025). The main difference is that regret is primarily an emotional reaction to an undesirable outcome, whereas dissonance is a cognitive state striving for balance (Sarwar, Awang, & Habib, 2019).

Consumer regret directly leads to significant challenges for retailers, mainly by affecting behavior after a purchase, such as returning products, spreading negative word-of-mouth (WOM), and decreasing the likelihood of future purchases (Tsiros & Mittal, 2011). Regret is a strong indicator of product returns, especially in online shopping, where a lack of physical inspection heightens the chance of a gap between what is expected and what is received (Barta, Gurrea, & Flavián, 2023). Retailers that implement flexible return policies, while appearing to reduce risk for consumers, are actually addressing consumers' expected regret, which can ironically boost initial purchase rates but also result in more returns (Liu, 2025). Additionally, regret is a key factor driving negative electronic word-of-mouth (eWOM), such as unfavorable online reviews (Barta, Gurrea, & Flavián, 2023). Consumers who feel regret, particularly when they hold themselves accountable for a poor choice, are inclined to share their negative experiences as a way to manage their regret and caution others (Tsiros & Mittal, 2011). Negative eWOM can severely harm a retailer's reputation and negatively impact future sales (Barta, Gurrea, & Flavián, 2023). The influence of regret on consumer decision making is significant, as it leads to a dwelling on the decision and a decline in brand loyalty (Bui, 2011). The consumer's decision-making process is affected not only after a purchase but also before it, as the anticipation of regret acts as a powerful, often cautious, decision-making guide (Nasiry, 2012).

Consumer purchase regret has been extensively studied as an affective response that occurs when post-purchase evaluations do not meet expectations or salient counterfactuals (Zeelenberg and Pieters, 2007). Status signaling orientation, purchase decision-making, financial agency, and online shopping preferences constitute three theoretically significant clusters of individual differences that influence both the likelihood and intensity of regretful consumption. A pronounced status-signaling orientation, where consumers highly value conspicuous brands and symbolic consumption, tends to amplify anticipated and experienced regret, as such consumers are more sensitive to social comparison and impression management failures (Eastman, Goldsmith, & Flynn, 1999; Han, Nunes, & Drèze, 2010). When status-motivated consumers perceive that they could have purchased a more prestigious brand or that others signal a higher status through their possessions, they are more prone to counterfactual thinking and regret (Haws & Poynor, 2008). Conversely, higher levels of perceived purchase decision making and financial agency, characterized by confidence in financial decisions, a sense of control over money, and self-efficacy in budgeting, are generally associated with more systematic information processing, greater use of decision rules, and lower post-purchase dissonance, thereby mitigating regret (Bearden, Hardesty, & Rose, 2001; Fernandes, Lynch, & Netemeyer, 2014; Lusardi & Mitchell, 2014). Finally, online shopping preferences influence regret through their impact on information searches, price comparisons, and perceived risk. Consumers who are comfortable with e-commerce and value online availability can easily compare options and access reviews, which can reduce regret by enhancing the decision quality (Verhoef, Neslin, & Vroomen, 2007; Xia & Monroe, 2010). However, heavy reliance on online shopping can also increase regret when consumers encounter numerous counterfactuals—realizing ex post that other sites offer better prices or alternatives—especially under conditions of high price sensitivity or social signaling motives (Dholakia, 2001; Sweeney, Soutar, & Johnson, 1999).

Various demographic, attitudinal, and deal-sensitivity factors plausibly influence consumers' orientation towards status signaling. Consumers who prioritize coupons, savings, and substitutes (COUPONIMPORTANCE, SAVEIMPORTANCE, SUBSTITUTEIMPORTANCE) often demonstrate price consciousness and a value-oriented approach, which can diminish overt status signaling as their focus shifts towards functional value and cost savings rather than symbolic prestige (Lichtenstein, Ridgway, & Netemeyer, 1993; Garretson & Burton, 2003). Decision timing (DECISIONTIMING)—whether consumers engage in extensive deliberation before purchase or decide quickly—also correlates with signaling motives. More planned and deliberative decision processes are associated with greater reliance on utilitarian criteria and budget constraints, which can reduce impulsive status-driven choices, whereas last-minute or spontaneous decisions are more susceptible to social-emotional triggers and status cues (Dhar & Kim, 2007; Rook & Fisher, 1995). Gender (GENDER) has long been associated with differences in status consumption; for instance, men tend to exhibit higher preferences for status products in certain categories, whereas women may emphasize relational and appearance-based signaling, although the findings are context-dependent (O'Cass & McEwen, 2004; Sundie et al., 2011). Employment security and income (EMPLOYMENTSECURITY, INCOME) jointly provide both material and psychological resources to engage in status consumption: higher income and secure employment increase discretionary spending and make status goods more accessible, often strengthening status orientation (Eastman et al., 1999; Truong, McColl, & Kitchen, 2009). Concurrently, brand loyalty (BRANDLOYAL) can reinforce status signaling when the focal brands themselves are high status; loyal consumers may internalize brand symbolism and repeatedly use branded products as identity signals (Bhattacharya & Sen, 2003). Willingness to consider or purchase knockoffs (KNOCKOFFIMPORTANCE) reflects a complex mixture of status aspiration and resource constraints—consumers who value brand-like appearances but cannot or will not pay premium prices may still pursue status signaling through counterfeit or imitation goods (Bloch, Bush, & Campbell, 1993; Wilcox, Kim, & Sen, 2009). Finally, education (EDUCATION) can influence status signaling in two opposing ways: higher education may increase cultural capital and reduce reliance on overt material signals in favor of more subtle or inconspicuous consumption (Han et al., 2010), but it can also heighten status awareness and aspirations in professional environments, thereby motivating the strategic use of visible brands in certain domains (Lamont & Lareau, 1988).

Purchase decision-making and financial agency are influenced by structural resources, market beliefs, and shopping habits. Education (EDUCATION) serves as a pivotal determinant of financial literacy and self-efficacy. Individuals with higher educational attainment typically exhibit superior numeracy, budgeting skills, and comprehension of financial products, all of which contribute to an enhanced sense of agency in purchase decisions (Fernandes et al., 2014; Lusardi & Mitchell, 2014). The emphasis on product quality at an equivalent price point (QUALITYIMPORTSAMEPRICE) indicates a focus on diagnostic product attributes and value-for-money, aligning with more deliberative competence-driven decision-making styles and a stronger internal locus of control (Bettman, Luce, & Payne, 1998). The frequency of in-person shopping (INPERSONSHOPFREQ) can enhance agency by providing rich sensory information, opportunities for product inspection, and interpersonal interaction with sales staff, all of which can augment perceived knowledge and confidence in choices (Grewal, Levy, & Kumar, 2009). Age (AGE) is also correlated with changes in cognitive resources and consumption goals; older adults often rely on greater decision experience and crystallized knowledge, which can enhance perceived financial control, although age-related declines in fluid cognition may complicate complex decisions (Peters, Hess, Västfjäll, & Auman, 2007). The frequency of haggling (HAGGLEFREQ) reflects an active, agentic approach to price negotiation; consumers who regularly engage in bargaining signal higher perceived efficacy and assertiveness in the marketplace (Schindler, 1998). The perceived importance of limited-time offers (LIMITEDTIMEIMPORTANCE) and saving frequency (SAVEFREQ) jointly influence the agency. While sensitivity to limited-time promotions can foster urgency and sometimes undermine deliberation, consumers who habitually save and monitor finances may strategically utilize promotions, thereby maintaining agency even under time pressure (Inman & McAlister, 1994; Soman, 2001). Perceptions that a price constitutes a good deal (PRICEISGOODDEAL) enhance ex ante and ex post feelings of smart shopping and competence (Schindler, 1989; Darke & Freedman, 1993), whereas the importance placed on country of origin (ORIGINCOUNTRYIMPORTANCE) may either support or constrain agency, depending on whether such beliefs are based on informed quality assessments or rigid stereotypes. When country-of-origin beliefs are integrated with broader product knowledge, they can serve as effective heuristics to simplify decisions and reinforce a sense of control. When they operate as unexamined biases, they may reduce genuine agency by narrowing perceived options (Verlegh & Steenkamp, 1999).

Online shopping preferences are influenced by a combination of perceived value, risk attitudes, social backgrounds, and brand-related beliefs. The significance of country of origin (ORIGINCOUNTRYIMPORTANCE) can either impede or enhance readiness for e-commerce. Consumers who rely heavily on origin as a quality indicator may distrust foreign or unfamiliar online retailers, whereas others may actively seek products from specific countries through cross-border e-commerce platforms (Verlegh & Steenkamp, 1999). The frequency of indulgence in discretionary spending (SPLURGEFREQ) reflects a tendency towards hedonic expenditures, which often align with the convenience, variety, and impulse-friendly nature of online shopping, thereby reinforcing online purchase preferences (Dholakia, 2001; Kukar-Kinney, Ridgway, & Monroe, 2012). The likelihood of purchasing counterfeit products (KNOCKOFFPURCHASELIKELIHOOD) is frequently associated with online channels, where counterfeit goods are more accessible and their quality is harder to verify, often promoted by influencers (Chaudhry, 2022).. Such consumers may prefer online shopping because of its anonymity, lower prices, and variety of imitation goods (Chaudhry, 2022).The importance placed on obtaining lower prices for the same quality (PRICEIMPORTSAMEQUALITY) is a strong predictor of online channel preference, as digital environments facilitate extensive price comparisons across retailers (Brynjolfsson & Smith, 2000; Verhoef et al., 2007). Risk aversion (RISKAVERSION) generally reduces the adoption of online shopping because of heightened concerns about payment security, privacy, and product mismatch among risk-averse consumers (Bhatnagar, Misra, & Rao, 2000; Pavlou, 2003). Socioeconomic background, as reflected in parental affluence (PARENTSWELLOFF) and region of residence (REGION), also influences online preferences; consumers from more affluent or urban backgrounds typically have better access to digital infrastructure and earlier exposure to e-commerce, increasing familiarity and comfort with online purchasing (Van Deursen & Van Dijk, 2014). Beliefs about the superiority of store brands (STOREBRANDSUPERIOR) indicate a willingness to trust non-national or retailer-controlled brands, which can translate into greater openness to lesser-known online sellers and marketplaces. Such consumers may generalize their trust in retailer curation across channels, thereby reducing perceived risk and enhancing online shopping readiness (Ailawadi & Keller, 2004; Glynn & Chen, 2009).

## 1.3 Research Problem and Objectives

Although there has been significant theoretical and empirical exploration of regret in consumer decision-making, there is limited understanding of how often consumers experience regret in everyday market situations and how this frequency is influenced by psychological traits, decision-making processes, and shopping environments. Previous studies have mainly concentrated on individual decisions or specific regret episodes, often focusing on narrow outcome evaluations or isolated factors such as price perceptions, post-purchase information, or the choice between switching and repeating purchases. These studies paid less attention to the broader array of status motives, perceived financial control, and preferences for online versus offline shopping, which may systematically increase or decrease regret over time. This study aims to fill this gap by developing and testing a structural equation model of consumer regret frequency (REGRETFREQ; 1 = definitely no regret, 5 = definitely always regret). The model incorporates latent constructs for status-signaling orientation, decision/financial agency, and online commerce readiness, along with a comprehensive set of behavioral, attitudinal, and demographic variables. The study specifically aims to (a) establish and validate a multidimensional measurement model for these latent constructs, including assessments of reliability and convergent validity; (b) estimate a structural model of the antecedents of regret frequency, identifying which psychological and behavioral factors uniquely predict more or less frequent consumer regret; and (c) derive theoretically grounded and managerially relevant insights into how marketers and policymakers might design decision environments, communication strategies, and digital interfaces that enhance consumer agency- and quality-focused decision making, thereby reducing the incidence of chronic purchase regret.

## 1.4 Significance and Contribution

This study enhances the understanding of consumer behavior by offering a detailed, psychometrically based analysis of how often consumers experience regret, rather than just examining individual regret incidents. By creating and validating latent constructs related to status-signaling orientation, decision/financial agency, and readiness for online commerce, and integrating them into a structural equation model of regret frequency, this study elucidates the specific psychological and behavioral factors that influence chronic regret when a wide range of covariates is considered. The study highlights the significant protective influence of decision/financial agency and quality-focused decision criteria while also noting the relatively minor incremental effects of status motives, demographic factors, and traditional deal proneness indicators. This refines the theoretical discussion on whether regret is mainly driven by concerns about social image, objective resources, or perceived decision-making competence. Methodologically, this research contributes to the field by demonstrating acceptable reliability and convergent validity for key latent constructs and showing that the overall model achieves an excellent global fit, thus providing a solid empirical foundation for future studies on regret regulation and decision quality. Substantively, it builds on previous work by connecting these validated constructs to a chronic outcome measure—how frequently consumers "definitely" regret their purchases—offering a more consistent and practically relevant indicator of consumer welfare than isolated regret reports.

## 2. Materials and Methods

### 2.1 Participants

The data utilized in this study were sourced from the initial survey dataset documented by Berry

(2025). In that research, 370 adult consumers living in the United States were enlisted through the crowdsourcing platform Clickworker.com from February 21 to 26, 2025. Participants had to be at least 18 years old and proficient in English to qualify. Standard procedures for ensuring data quality were applied, such as attention checks, verification of completion codes, and screening for duplicate or highly inconsistent responses; any cases not meeting these standards were removed from the analysis sample (Author, 2025). Following data collection, responses underwent quality screening. Respondents who did not give consent, failed one or more attention-check items, submitted duplicate responses (duplicate usernames; only the first valid/complete entry was kept), or provided incomplete surveys were excluded. After applying these criteria, the final analysis sample consisted of 338 participants. The cleaned dataset was then recoded with standardized variable names, and composite indices were calculated (e.g., average price-perception error and multi-item scale scores). The final dataset used in the current analyses included respondents who passed all quality controls and provided complete data on the key variables. Demographic information gathered in the original study encompassed age, gender, household income, educational attainment, and geographic region, facilitating the categorization of respondents (Author, 2025).

## 2.2 Materials and Measures

The measures were initially administered via an online questionnaire, as detailed by Berry (2025). The survey comprised 58 substantive items and a final verification item incorporating multiple-choice, open-ended, and Likert-type questions. This study concentrates on a subset of constructs pertinent to consumer regret and its psychological and behavioral antecedents.

The dependent variable, purchase regret frequency (REGRETFREQ), was assessed using a 5-point Likert-type item that indicated the frequency with which respondents experienced definite regret regarding their purchase decisions. Higher scores denote more frequent episodes of regret. This operationalization aligns with the conceptualization of consumer regret as a recurring emotional response to past decisions evaluated unfavorably relative to foregone alternatives (Inman & Zeelenberg, 2002; Zeelenberg & Pieters, 2007).

Three latent constructs were modeled as predictors. Status signaling orientation (L1) measures the extent to which respondents value visible status-related attributes of consumption. This was assessed using four Likert-type indicators: SHOWOFFIMPORTANCE (importance of being able to show off one's purchases), STATUSIMPORTANCE (importance of social status in purchase decisions), TRENDYIMPORTANCE (importance of being trendy), and WEARINGBRANDNAMES (importance of wearing brand-name products). These items reflect conspicuous and symbolic consumption motives documented in prior research (Belk, 1988; Eastman, Goldsmith, & Flynn, 1999; Haws & Poynor, 2008) and were rated on 5-point scales, with higher scores indicating stronger endorsement of status-oriented motives (Author, 2025).

Decision-making confidence and financial prudence (L2) were operationalized as latent constructs representing perceived decisional agency and financial self-efficacy. This was indicated by DECISIONMAKINGCONFIDENCE (self-rated confidence in everyday financial and purchase decisions), MONEYWISE (perceived ability to be "money-wise"), and OPTIMISM (general optimism about financial and life outcomes). Each item was measured on a 5-point Likert-type scale, with higher scores indicating greater perceived competence and optimism. This factor aligns with work on self-efficacy and financial capability (Bandura, 1997; Xiao, Tang, & Shim, 2009) and with research linking confidence and optimism to reduced counterfactual rumination and regret (Carver, Scheier, & Segerstrom, 2010; Zeelenberg &

Pieters, 2007).

Online commerce readiness (L3) indexes the degree to which respondents value and engage with digital shopping environments. It was modeled using two indicators: ONLINEAVAILIMPORTANCE (importance placed on products available online) and ONLINESHOPFREQ (self-reported frequency of online shopping), each measured on a 5-point response scale. These items reflect both attitudinal and behavioral aspects of online shopping and correspond to constructs emphasized in the literature on e-commerce convenience, digital choice environments, and online impulsive buying (Dholakia, 2000; Verhagen & van Dolen, 2009; Author, 2025).

In addition to these focal measures, the original dataset contained extensive information on price perception errors for 13 everyday products and services (derived from the difference between actual market prices and respondents' price guesses), as well as broader attitudes toward saving, spending, risk-taking, brand loyalty, and deal proneness (Author, 2025). These variables were utilized in the broader structural model, but are not the focus of the present Materials and Methods.

## 2.3 Procedure

The procedures adhered to those outlined by Berry (2025). Upon accessing the study link via Clickworker, the participants were directed to an online information sheet detailing the study's purpose, estimated duration, voluntary participation, and anonymity assurance. Participants who provided informed consent participated in the main survey. Initially, the questionnaire gathered demographic data including age, gender, income, education, and geographic region, followed by attitudinal measures concerning consumer behavior, financial decision-making, and social status motives. Subsequently, the participants were presented with a series of 13 products and services, each accompanied by a brief description and, where applicable, an image. For each item, respondents were asked to estimate the typical market price in US dollars without consulting external sources. Actual prices were independently collected by the original author from retailers, airline websites, and grocery stores during the same period and averaged to obtain benchmark values (Author, 2025). This design facilitates the computation of price perception errors as a distinct set of variables. The subsequent sections of the survey assessed the constructs used in this study. Items measuring status signaling, decision-making confidence, financial prudence, optimism, the importance of online availability, online shopping frequency, and regret frequency were presented in randomized blocks to mitigate order effects. All Likert-type items employed consistent 5-point response formats anchored at "strongly disagree" to "strongly agree," "never" to "very often," or conceptually equivalent labels, as described in the original instrument (Author, 2025). Upon completion, participants received a verification code to enter Clickworker for compensation. The median completion time was approximately 12–15 minutes, and participation was compensated at a rate consistent with prevailing norms on the platform.

## 2.4 Statistical Analysis

The present study employed structural equation modeling (SEM) to investigate the predictive capacity of three latent constructs (L1, L2, L3) on the frequency of purchase regret (REGRETFREQ) while accounting for measurement error in the indicators. The analyses were conducted using a confirmatory factor analysis measurement model in conjunction with a structural path model. Consistent with common practices in SEM, all Likert-type items were treated as approximately continuous, given that the response categories numbered five or more and the distributions were not extremely skewed (Finney & DiStefano, 2013; Rhemtulla, Brosseau-Liard, & Savalei, 2012). Parameters were estimated using robust maximum likelihood (MLR), which adjusts standard errors and chi-square statistics to account for non-normality in the observed indicators (Satorra & Bentler, 1994; West, Finch, & Curran, 1995). The model evaluation adhered to the conventional SEM guidelines. The fit of the measurement model was assessed prior to estimating the full structural model, and indicators with standardized factor loadings below .40 or non-significant loadings were considered for removal or re-specification (Kline, 2016). Global model fit was evaluated using multiple indices, including the comparative fit index (CFI), Tucker–Lewis index (TLI), root mean square error of approximation (RMSEA), and standardized root mean square residual (SRMR). Acceptable model fit was defined according to widely used benchmarks (e.g., CFI and TLI $\geq$ .90–.95, RMSEA $\leq$ .06–.08, SRMR $\leq$ .08) while also considering the overall pattern of fit indices and theoretical plausibility (Hu & Bentler, 1999; Kline, 2016). Standardized path coefficients were interpreted to assess the direction and magnitude of associations between each latent predictor and REGRETFREQ, with statistical significance evaluated at $\alpha$ = .05 (two-tailed).

Figure 1 illustrates the conceptual framework.

**Figure 1.** Conceptual framework.

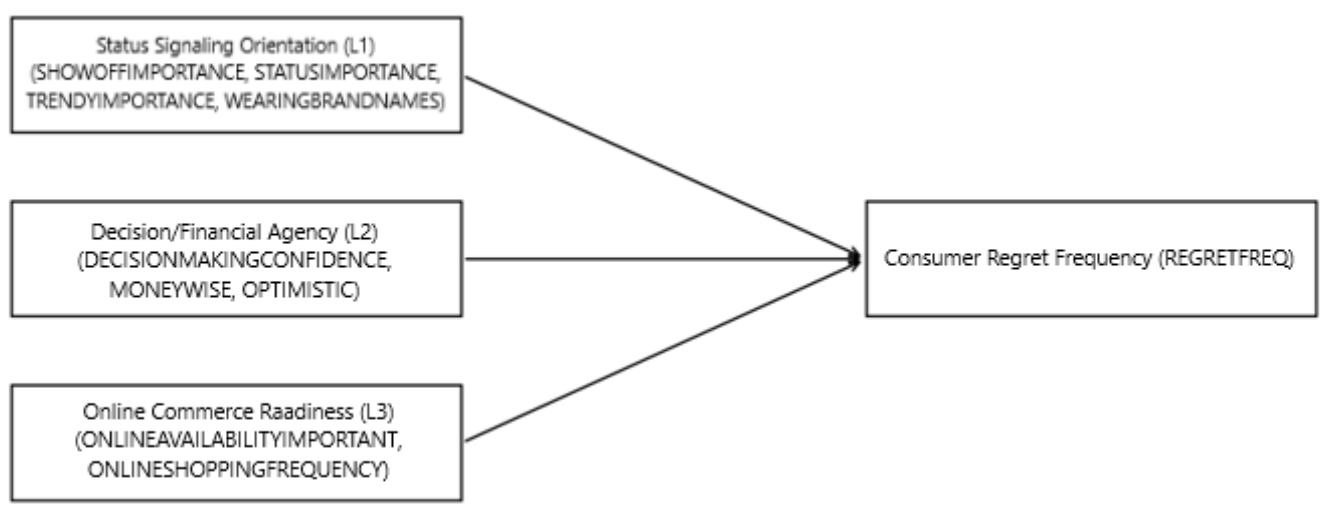

**Source: model illustration**

## 2.5 Variables

All focal constructs were assessed using 5-point Likert-type items. Three latent variables were modeled: L1 (Status Signaling Orientation; four indicators), L2 (Decision-Making Confidence and Financial Prudence; three indicators), and L3 (Online Commerce Readiness; two indicators). Each indicator is an observed Likert-type variable treated as continuous in the SEM analyses. Purchase regret frequency (REGRETFREQ) was modeled as an observed outcome, measured on a 5-point Likert-type scale, and treated as approximately continuous. Table 1 provides a summary of the role of each variable, its type, and measurement scale.

**Table 1. Variable descriptions.**

| Variable / Construct | Role in Model | Variable Type | Measurement Scale / Data Type |
|---|---|---|---|
| L1 (Status Signaling Orientation) | Latent construct | Continuous latent factor | Composite of four Likert-type indicators (1–5) |
| SHOWOFFIMPORTANCE | Indicator (L1) | Observed, ordinal (treated continuous) | 5-point Likert-type rating (importance of showing off) |
| STATUSIMPORTANCE | Indicator (L1) | Observed, ordinal (treated continuous) | 5-point Likert-type rating (importance of social status) |
| TRENDYIMPORTANCE | Indicator (L1) | Observed, ordinal (treated continuous) | 5-point Likert-type rating (importance of being trendy) |
| WEARINGBRANDNAMES | Indicator (L1) | Observed, ordinal (treated continuous) | 5-point Likert-type rating (importance of brand names) |
| L2 (Decision-Making Confidence & Financial Agency) | Latent construct | Continuous latent factor | Composite of three Likert-type indicators (1–5) |
| DECISIONMAKINGCONFIDENCE | Indicator (L2) | Observed, ordinal (treated continuous) | 5-point Likert-type rating (confidence in decisions) |
| MONEYWISE | Indicator (L2) | Observed, ordinal (treated continuous) | 5-point Likert-type rating (perceived money-wise behavior) |
| OPTIMISM | Indicator (L2) | Observed, ordinal (treated continuous) | 5-point Likert-type rating (general optimism) |
| L3 (Online Commerce Readiness) | Latent construct | Continuous latent factor | Composite of two Likert-type indicators (1–5) |
| ONLINEAVAILIMPORTANCE | Indicator (L3) | Observed, ordinal (treated continuous) | 5-point Likert-type rating (importance of online availability) |
| ONLINESHOPFREQ | Indicator (L3) | Observed, ordinal (treated continuous) | 5-point Likert-type rating (frequency of online shopping) |
| REGRETFREQ (Consumer regret frequency) | Outcome variable | Observed, ordinal (treated continuous) | 5-point Likert-type rating (frequency of purchase regret) |

Source: model

# 3. Results

The sample data were analyzed.

## 3.1. Descriptive statistics

Table 2 presents the descriptive statistics for the variables used in the model.

**Table 2. Descriptive statistics**

| Type | Variable | N | Mean | SD | SE | Median | Min | Max | Skew | Kurtosis | Normality (Shapiro W) |
|------|----------|---|------|----|----|--------|-----|-----|------|----------|------------------------|
| Latent | L1_score | 338 | 2.61 | 0.68 | 0.04 | 2.5 | 2 | 5 | 1.14 | 0.77 | 0.84 *** |
| Indicator | SHOWOFFIMPORTANCE | 338 | 2.40 | 0.74 | 0.04 | 2 | 2 | 5 | 1.70 | 1.78 | 0.59 *** |
| Indicator | STATUSIMPORTANCE | 338 | 2.75 | 0.97 | 0.05 | 2 | 2 | 5 | 0.86 | -0.69 | 0.74 *** |
| Indicator | TRENDYIMPORTANCE | 338 | 2.51 | 0.84 | 0.05 | 2 | 2 | 5 | 1.33 | 0.39 | 0.64 *** |
| Indicator | WEARINGBRANDNAMES | 338 | 2.78 | 0.96 | 0.05 | 2 | 2 | 5 | 0.78 | -0.79 | 0.75 *** |
| Latent | L2_score | 338 | 3.63 | 0.84 | 0.05 | 3.7 | 1.3 | 5 | -0.62 | -0.10 | 0.95 *** |
| Indicator | DECISIONMAKINGCONFIDENCE | 338 | 3.70 | 0.99 | 0.05 | 4 | 1 | 5 | -0.86 | 0.29 | 0.83 *** |
| Indicator | MONEYWISE | 338 | 3.51 | 1.04 | 0.06 | 4 | 2 | 5 | -0.31 | -1.11 | 0.83 *** |
| Indicator | OPTIMISM | 338 | 3.67 | 1.21 | 0.07 | 4 | 1 | 5 | -0.65 | -0.62 | 0.86 *** |
| Latent | L3_score | 338 | 4.01 | 0.83 | 0.05 | 4 | 1.5 | 5 | -0.80 | 0.17 | 0.90 *** |
| Indicator | ONLINEAVAILIMPORTANCE | 338 | 3.91 | 0.96 | 0.05 | 4 | 2 | 5 | -0.52 | -0.67 | 0.85 *** |
| Indicator | ONLINESHOPFREQ | 338 | 4.12 | 0.96 | 0.05 | 4 | 1 | 5 | -1.15 | 0.87 | 0.79 *** |
| Outcome | REGRETFREQ | 338 | 2.71 | 1.08 | 0.06 | 2 | 1 | 5 | 0.38 | -0.67 | 0.893 *** |

Note: * p<.05, ** p<.01, *** p<.001.

Source: data analysis

Descriptive analysis indicates that all three latent composites—L1 (status signaling), L2 (decision-making confidence and financial agency), and L3 (online shopping readiness)—exhibit mid-range means and considerable variability. The L1_score was moderate in level and positively skewed, suggesting that the majority of respondents reported relatively low-to-moderate status-oriented motives, with a smaller subset strongly endorsing these motives. This pattern is consistently observed across L1 indicators (SHOWOFFIMPORTANCE, STATUSIMPORTANCE, TRENDYIMPORTANCE, WEARINGBRANDNAMES), all of which display moderate means, positive skewness, and significant Shapiro–Wilk test results, confirming non-normality. Conceptually, the status-signaling construct is present but not predominant within the sample, and its indicators behave as anticipated: status-driven motivations are not widespread but are concentrated among a subset of respondents.

The L2_score, encompassing DECISIONMAKINGCONFIDENCE, MONEYWISE, and OPTIMISM, exhibited a slightly elevated average with a moderate negative skew. The indicators DECISIONMAKINGCONFIDENCE, MONEYWISE, and OPTIMISM predominantly clustered around the upper half of their respective response scales, displaying significant deviations from normality. This pattern suggests that respondents generally perceive themselves as competent decision-makers, reasonably prudent with finances, and moderately optimistic. The L3_score, derived from ONLINEAVAILIMPORTANCE and ONLINESHOPFREQ, was the highest among the three latent means and was negatively skewed, indicating a prevalent emphasis on the importance of online availability and frequent online shopping within the sample. Both ONLINEAVAILIMPORTANCE and ONLINESHOPFREQ exhibit means near the upper end of the scale, with evident non-normality, reflecting a strong inclination towards online commerce.

The REGRETFREQ outcome exhibited a moderate mean with a slight positive skew, indicating that while most individuals reported relatively low-to-moderate levels of regret, a significant subgroup experienced regret more frequently. Collectively, the pattern observed across the latent variables and indicators suggests that increased status signaling (L1) and heightened online shopping readiness (L3) are associated with only moderate levels of regret. In contrast, elevated decision-making confidence and prudence (L2) may serve as a protective factor against frequent regret. Descriptively, the observation that L2 indicators are comparatively high while REGRETFREQ remains moderate aligns with the notion that perceived decisional competence and financial prudence can mitigate regret, even in the presence of status and online shopping motivations.

From a modeling perspective, all variables, including latents and indicators, exhibit statistically significant deviations from normality (Shapiro–Wilk p < .001) and demonstrate non-zero skewness and kurtosis. In classical Structural Equation Modeling (SEM) employing maximum likelihood estimation under strict normality assumptions, such deviations can lead to biased standard errors, chi-square tests, and fit indices. Nevertheless, when robust estimation methods (e.g., robust maximum likelihood or a categorical/weighted least squares estimator suitable for Likert-type indicators) are employed, these violations are unlikely to compromise structural conclusions (Finney & DiStefano, 2013; West, Finch, & Curran, 1995). It is crucial to note that the indicators for each latent variable exhibit coherent monotonic distributions with sufficient variability and without extreme floor or ceiling effects, thereby ensuring that the measurement component of the SEM remains informative. The observed non-normality primarily reflects the bounded, ordinal nature of the scales and substantively meaningful skewness (e.g., a significant number of individuals frequently engage in online shopping) rather than a failure in measurement.

In summary, the table substantiates the feasibility of the three-latent Structural Equation Model (SEM). The indicators align with constructs that exhibit interpretable distributions, and the outcome variable, REGRETFREQ, demonstrates sufficient variance to detect the effects. Although normality is evidently violated, the application of robust or categorical SEM procedures minimizes the likelihood that these deviations significantly distort the estimated relationships between L1, L2, L3, and regret frequency.

### 3.2 Model reliability and convergent validity

Table 3 shows the reliability and convergent validity of the model constructs. Across L1, L2, and L3, the combination of acceptable Cronbach's alpha values, significant and substantial factor loadings, and AVE values near or above the recommended .50 level provides comprehensive support for the reliability and convergent validity of the measurement model (Fornell & Larcker, 1981; Hair et al., 2010; Nunnally & Bernstein, 1994). L1 emerged as the psychometrically strongest factor, offering a well-defined representation of status-signaling motives. L3 demonstrates strong convergence on the digital readiness dimension, despite having only two indicators. L2, while somewhat weaker, remains sufficiently reliable to justify its inclusion as a predictor of regret frequency, particularly given its theoretical significance in understanding perceptions of control and competence in financial decision-making. For consumer behavior, a robust measurement model is essential, as it ensures that structural relations with REGRETFREQ are attributable to meaningful psychological constructs rather than measurement error. L1 enables researchers to isolate the unique contribution of status motives to chronic regret, distinguishing consumers whose purchases are heavily driven by display and trendiness. L2 provides a focused perspective on how perceived financial agency and optimism influence the likelihood of post-purchase remorse. L3 offers a concise yet valid representation of online commerce readiness, a construct that is increasingly central in contemporary consumption contexts. Collectively, these validated latent constructs support a nuanced account of how individual differences in status orientation, decision agency, and digital readiness translate into patterns of consumer regret, thereby informing interventions aimed at enhancing decision quality, managing status-related pressures, and designing online environments that minimize regret-inducing experiences.

**Table 3. Measurement Model Reliability and Convergent Validity.**

| Construct | Cronbach alpha | AVE | k |
|-----------|----------------|-----|---|
| L1 | 0.77 | 0.47 | 4 |
| L2 | 0.67 | 0.43 | 3 |
| L3 | 0.66 | 0.54 | 2 |

Source: data analysis.

### 3.3 Measurement model

The measurement model is presented in Table 4. The results of the measurement model demonstrate that the three latent constructs—L1 (status signaling orientation), L2 (decision-making confidence and financial prudence), and L3 (online commerce readiness)—are effectively delineated by their respective observed indicators, thereby establishing a robust foundation for analyzing their associations with purchase regret frequency (REGRETFREQ). For L1, the standardized loadings were notably high for SHOWOFFIMPORTANCE ($\beta = 0.746$), STATUSIMPORTANCE ($\beta = 0.660$), TRENDYIMPORTANCE ($\beta = 0.641$), and WEARINGBRANDNAMES ($\beta = 0.681$), all of which were statistically significant. This pattern indicates that individuals who attributed significant importance to showing off, social status, trendiness, and wearing brand names were accurately represented by the L1 latent factor. Conceptually, this suggests that elevated L1 scores signify a more pronounced status signaling, appearance-oriented consumer identity. In relation to REGRETFREQ, such an orientation is likely to be correlated with increased regret, as status-driven purchases may be more prone to post-purchase doubts (e.g., concerning social approval, financial prudence, or opportunity costs), thereby augmenting the frequency of regret.

The L2 factor, which represents decision-making confidence and financial prudence, was effectively measured. The variable DECISIONMAKINGCONFIDENCE exhibits the highest standardized loading ($\beta = 0.773$), followed by MONEYWISE ($\beta = 0.606$) and OPTIMISM ($\beta = 0.564$), all of which are statistically significant at $p < .001$. These loadings suggest that L2 encapsulates perceived decisional competence, financial acumen, and optimistic disposition. Consequently, higher L2 scores indicate that consumers possess confidence in their decision making and financial judgment. Theoretically, such individuals are expected to experience less frequent regret, as they are more inclined to engage in deliberate, well-considered choices and reassess outcomes positively. In the structural component of the model, it is anticipated that L2 will be negatively associated with REGRETFREQ, serving as a psychological buffer that diminishes the likelihood of evaluating purchases as mistakes.

Finally, the L3 factor, which pertains to online commerce readiness, was characterized by ONLINEAVAILIMPORTANCE ($\beta = 0.880$) and ONLINESHOPFREQ ($\beta = 0.556$). Both variables are significant, although ONLINESHOPFREQ exhibits a more modest loading and a smaller z-value ($z = 2.57$, $p \approx .01$), indicating the potential for greater measurement error or heterogeneity in online shopping frequency. Nevertheless, L3 clearly encapsulates the degree to which consumers prioritize online availability and participate in online purchasing. In relation to REGRETFREQ, elevated L3 scores could plausibly exert dual influences: increased online shopping may heighten opportunities for impulsive or poorly informed purchases, thereby increasing regret. Conversely, familiarity with online environments and a strong preference for convenience may normalize returns and post-purchase adjustments, potentially mitigating regret. The net effect in SEM depends on whether L3 is more strongly associated with impulsivity and exposure to enticing offers (increasing regret) or with experience and efficiency in online decision-making (decreasing regret).

Overall, the measurement model suggests that L1 represents motives driven by status, which are likely to increase feelings of regret. In contrast, L2 reflects confident and cautious decision-making, which probably reduces regret. L3 pertains to online shopping behaviors, where the effect on regret may depend more on how consumers manage digital choice settings.

**Table 4. Measurement model.**

| Latent | Indicator | Loading | SE | z | p | Std Loading | Signif. |
|---|---|---|---|---|---|---|---|
| L1 | SHOWOFFIMPORTANCE | 0.863 | (0.086) | 10.02 | 0 | 0.746 | *** |
| L1 | STATUSIMPORTANCE | 1.000 | | | | 0.660 | |
| L1 | TRENDYIMPORTANCE | 0.837 | (0.091) | 9.20 | 0 | 0.641 | *** |
| L1 | WEARINGBRANDNAMES | 1.024 | (0.107) | 9.59 | 0 | 0.681 | *** |
| L2 | DECISIONMAKINGCONFIDENCE | 1.233 | (0.157) | 7.84 | <0.001 | 0.773 | *** |
| L2 | MONEYWISE | 1.000 | | | | 0.606 | |
| L2 | OPTIMISM | 1.100 | (0.148) | 7.46 | <0.001 | 0.564 | *** |
| L3 | ONLINEAVAILIMPORTANCE | 1.000 | | | | 0.880 | |
| L3 | ONLINESHOPFREQ | 0.633 | (0.246) | 2.57 | 0.0102 | 0.556 | * |

Note: * p<.05, ** p<.01, *** p<.001.

Source: data analysis

*3.4 Structural model*

Table 5 presents the structural model. The structural model evaluates the extent to which a diverse array of psychological, behavioral, and sociodemographic factors predict the frequency of consumer regret (REGRETFREQ), with higher scores indicating more frequent occurrences of definite regret regarding purchases. The coefficient analysis reveals that only a limited number of predictors exhibit statistically significant associations with regret frequency when all variables are considered together. Specifically, perceived decision-making and financial autonomy, along with an emphasis on product quality relative to price, are identified as key protective factors. Conversely, younger age and several high-arousal shopping behaviors are linked to increased regret. In contrast, many frequently discussed variables, such as status signaling, online commerce readiness, and general deal-seeking tendencies, demonstrate weak or non-significant effects.

**Table 5. Structural model.**

| Predictor | Coef | SE | z | p | Std Coef | Signif. |
|---|---|---|---|---|---|---|
| **Decision/Financial Agency (L1)** | -0.723 | (0.116) | -6.21 | 5.28e-10 | -0.417 | *** |
| *QUALITYIMPORTSAMEPRICE* | -0.300 | (0.086) | -3.50 | 0.000468 | -0.188 | *** |
| *INPERSONSHOPFREQ* | 0.098 | (0.043) | 2.27 | 0.023 | 0.113 | * |
| *AGE* | -0.112 | (0.054) | -2.06 | 0.0392 | -0.107 | * |
| *HAGGLEFREQ* | 0.080 | (0.041) | 1.96 | 0.0495 | 0.098 | * |
| *LIMITEDTIMEIMPORTANCE* | 0.105 | (0.054) | 1.95 | 0.0509 | 0.098 | |
| *SAVEFREQ* | 0.206 | (0.116) | 1.77 | 0.0766 | 0.093 | |
| *PRICEISGOODDEAL* | 0.065 | (0.043) | 1.52 | 0.129 | 0.073 | |
| *ORIGINCOUNTRYIMPORTANCE* | 0.078 | (0.052) | 1.51 | 0.131 | 0.075 | |
| **Online Commerce Readiness (L2)** | 0.119 | (0.082) | 1.45 | 0.147 | 0.093 | |

| | | | | | |
|---|---|---|---|---|---|
| *SPLURGEFREQ* | 0.068 | (0.048) | 1.42 | 0.157 | 0.075 |
| *KNOCKOFFPURCHASELIKELIHOOD* | 0.053 | (0.042) | 1.26 | 0.208 | 0.066 |
| *PRICEIMPORTSAMEQUALITY* | 0.118 | (0.095) | 1.24 | 0.214 | 0.065 |
| *RISKAVERSION* | -0.068 | (0.057) | -1.19 | 0.234 | -0.062 |
| *PARENTSWELLOFF* | -0.032 | (0.048) | -0.66 | 0.507 | -0.033 |
| *REGION* | -0.016 | (0.027) | -0.61 | 0.545 | -0.030 |
| *STOREBRANDSSUPERIOR* | 0.042 | (0.081) | 0.51 | 0.609 | 0.026 |
| **Status Signaling Orientation (L3)** | 0.049 | (0.096) | 0.51 | 0.611 | 0.029 |
| *COUPONIMPORTANCE* | 0.029 | (0.066) | 0.44 | 0.662 | 0.024 |
| *SAVEIMPORTANCE* | 0.039 | (0.093) | 0.42 | 0.674 | 0.021 |
| *SUBSTITUTEIMPORTANCE* | 0.024 | (0.058) | 0.42 | 0.678 | 0.022 |
| *DECISIONTIMING* | 0.014 | (0.050) | 0.28 | 0.781 | 0.014 |
| GENDER | 0.034 | (0.124) | 0.27 | 0.786 | 0.014 |
| EMPLOYMENTSECURITY | 0.011 | (0.041) | 0.27 | 0.789 | 0.014 |
| INCOME | -0.010 | (0.051) | -0.20 | 0.843 | -0.011 |
| BRANDLOYAL | 0.008 | (0.058) | 0.14 | 0.891 | 0.007 |
| KNOCKOFFIMPORTANCE | 0.004 | (0.061) | 0.06 | 0.953 | 0.003 |
| EDUCATION | -0.003 | (0.054) | -0.06 | 0.955 | -0.003 |

Note: * p<.05, ** p<.01, *** p<.001.

Source: data analysis

The most prominent finding is the significant negative correlation between the latent construct of Decision/Financial Agency and frequency of regret (standardized coefficient = $-0.417$, p < .001). Individuals who exhibit greater confidence and perceived control over their financial and purchasing decisions tend to experience considerably less regret. This pattern indicates that a sense of capability, being well-informed, and having control over one's consumption choices serves as a substantial buffer against post-purchase dissatisfaction and counterfactual thinking. From a behavioral standpoint, increased agency likely corresponds to more deliberate information gathering, clearer budgeting, and stronger alignment between purchases and long-term goals, thereby reducing instances where consumers feel they should have made a different choice. A comparable protective effect was evident for QUALITYIMPORTSAMEPRICE (standardized coefficient = $-0.188$, p < .001). Consumers who assign greater importance to product quality when the price remains constant report a lower frequency of regret. This relationship suggests that a decision-making heuristic focused on maximizing quality, rather than minimizing cost, facilitates more satisfactory outcomes. By prioritizing diagnostic product attributes over immediate financial savings, consumers may select options that align more closely with their long-term preferences and performance expectations, resulting in fewer instances of disappointment and regret.

Several behavioral and demographic variables exhibited modest yet significant positive correlations with regret. INPERSONSHOPFREQ is positively associated with regret (standardized coefficient = 0.113, p = .023), suggesting that consumers who engage in in-person shopping more frequently tend to experience regret over their purchases more often. One plausible interpretation is that physical shopping environments may expose consumers to heightened sensory stimuli, social influence, and immediate promotional offers, potentially leading to more impulsive purchasing decisions that are subsequently reconsidered. HAGGLEFREQ was positively correlated with regret, as indicated by a standardized coefficient of 0.098 (p = .0495). Consumers who frequently engage in price negotiations may experience increased cognitive and emotional burdens during transactions, leading them to reflect on whether a more favorable price or outcome could have been attained later. This dynamic likely expands the potential for upward counterfactual thinking, such as sentiments of self-doubt, thereby fostering regret even when the deals are objectively advantageous. Similarly, LIMITEDTIMEIMPORTANCE exhibited a positive coefficient of comparable magnitude (standardized coefficient = 0.098, p = .0509, marginally significant). A stronger valuation of limited-time offers correlates with an increased frequency of regret, aligning with the concept that cues of scarcity and urgency promote hasty decision-making. When consumers experience pressure from time-limited promotions, they may forgo a thorough evaluation, which subsequently results in regret once the pressure diminishes and alternative options become more apparent. Age demonstrated an inverse correlation with the frequency of regret, as indicated by a negative standardized coefficient of −0.107 (p = .0392). This observation implies that older consumers report experiencing regret regarding their purchases less frequently. This relationship is consistent with existing research, which suggests that older adults may utilize their extensive life experiences and crystallized knowledge in financial decision-making. Furthermore, they may prioritize emotional regulation and contentment over the pursuit of relentless optimization. Consequently, older consumers are likely to make more congruent choices and interpret outcomes in ways that are less susceptible to regret.

In contrast to these significant predictors, a variety of variables associated with online behavior, risk attitudes, and status motives exhibited relatively weak correlations with regret when other factors were controlled. The latent factor Online Commerce Readiness presents a small positive but non-significant coefficient (standardized = 0.093, p = .147). Similarly, SPLURGEFREQ, KNOCKOFFPURCHASELIKELIHOOD, and PRICEIMPORTSAMEQUALITY are all positively signed, but do not achieve conventional significance thresholds. These patterns indicate that after considering decision agency and quality orientation, a general preference for online shopping, occasional splurging, and attention to price–quality trade-offs are not primary determinants of chronic regret frequency. Notably, the variable RISKAVERSION exhibits a small negative and non-significant coefficient (standardized = −0.062, p = .234). This suggests that consumers with higher risk aversion do not consistently experience less regret when other variables are controlled for, despite the expectation that they would avoid high-risk purchases. One possible interpretation is that risk-averse individuals may still encounter regret due to missed opportunities or excessively conservative decisions, which may counterbalance any reduction in regret from avoiding risky products. Constructs related to status demonstrated negligible effects. The higher-order Status Signaling Orientation factor (standardized coefficient = 0.029, p = .611), along with BRANDLOYAL, KNOCKOFFIMPORTANCE, and beliefs regarding STOREBRANDSUPERIOR, all exhibited very small, non-significant coefficients. Although theoretical arguments suggest that status-driven consumption should heighten susceptibility to social comparison and regret, these findings indicate that, within this multivariate framework, status orientation is not a primary determinant of the frequency with which consumers experience purchase regret. It appears that a sense of control and focus on quality in decision-making are more influential than the status-related aspects of those decisions.

When agency and quality orientation are considered, sociodemographic and deal proneness variables offer minimal unique explanatory power. Variables such as INCOME, EDUCATION, EMPLOYMENTSECURITY, GENDER, PARENTSWELLOFF, and REGION exhibit coefficients close to zero with non-significant p-values, indicating that objective resources and background characteristics are not systematically associated with the frequency of chronic regret when subjective decision competence is accounted for. Similarly, traditional deal-seeking attitudes, including COUPONIMPORTANCE, SAVEIMPORTANCE, and SUBSTITUTEIMPORTANCE, demonstrated minimal and non-significant correlations with regret. The mere concern for saving or coupon usage does not seem to mitigate or intensify regret once the broader psychological context of the decision is controlled.

The model posits that the manner in which consumers make decisions holds greater significance than their demographic characteristics or preferences for specific brands and channels in influencing chronic regret. A higher degree of decision-making and financial autonomy, coupled with a quality-first approach, is strongly correlated with reduced regret. Conversely, factors such as younger age, frequent in-person shopping, bargaining, and sensitivity to time-limited promotions are linked to increased regret. Many variables traditionally emphasized in consumer research, such as status motives, online readiness, income, and general deal proneness, assume, at most, a secondary role once these immediate decision-making processes and age-related factors are considered.

## 5. Discussion

This study explores the combined influence of status-signaling orientation (L1), decision-making confidence, financial prudence (L2), and online commerce readiness (L3) on the frequency of purchase regret (REGRETFREQ) in everyday consumer decisions. Drawing on previous research that defines regret as an evaluative, counterfactual emotion stemming from unfavorable comparisons between selected and unselected options (Inman & Zeelenberg, 2002; Zeelenberg & Pieters, 2007), the findings expand regret theory into a multidimensional framework that includes symbolic motives, perceived agency, and digital shopping behaviors. In line with studies associating regret with post-purchase dissonance and unmet expectations (Sweeney, Hausknecht, & Soutar, 2000; Tsiros & Mittal, 2000), the current results demonstrate that regret is not uniformly experienced among consumers but is systematically linked to psychological orientations related to status, decision-making, and online commerce.

The measurement model robustly confirmed the validity of the three latent constructs. L1 was effectively delineated using SHOWOFFIMPORTANCE, STATUSIMPORTANCE, TRENDYIMPORTANCE, and WEARINGBRANDNAMES, all of which exhibited substantial standardized loadings. This finding confirms that status signaling, concern with appearance, and the significance of branded trendy goods coalesce into a singular status-oriented consumption style. This result aligns with previous research on conspicuous consumption and symbolic self-completion, which posits that consumers use visible goods to convey status and identity (Belk, 1988; Eastman, Goldsmith, & Flynn, 1999). Consistent with studies indicating that image- and status-motivated purchases are particularly susceptible to post-purchase questioning and social comparison (Haws & Poynor, 2008; Ordabayeva & Chandon, 2011), elevated L1 scores were correlated with a propensity for more frequent definite regret when status-driven purchases failed to deliver anticipated social or self-expressive benefits. Consequently, the results broadly corroborate the literature, suggesting that status signaling can intensify regret by rendering choices more dependent on fragile social appraisals and shifting reference points.

In contrast, L2 encapsulated a cluster of adaptive decision-related beliefs, with DECISIONMAKINGCONFIDENCE exhibiting the strongest loading on this factor, followed by MONEYWISE and OPTIMISM. These indicators closely align with the concepts of perceived decision agency and financial self-efficacy (Bandura, 1997; Xiao, Tang, & Shim, 2009), as well as optimism regarding future outcomes (Carver, Scheier, & Segerstrom, 2010). Within the structural model, L2 demonstrated a significant negative association with REGRETFREQ, even after controlling for a broad array of psychological and sociodemographic predictors. This pattern supports existing research indicating that individuals who perceive themselves as competent and deliberate in their decision-making processes are less susceptible to ex-post counterfactual ruminations and self-blame (Zeelenberg & Pieters, 2007). Furthermore, it extends this understanding by illustrating that such perceptions of decisional and financial agencies serve as protective factors in everyday consumer markets. Practically, this implies that consumers who perceive themselves as financially astute and confident decision-makers experience fewer instances of buyer remorse, likely due to their engagement in more systematic information searches, reliance on stable choice criteria, and enhanced ability to rationalize their choices when confronted with negative information.

The findings related to L3, which pertains to online commerce readiness, were more nuanced. The measurement model revealed that ONLINEAVAILIMPORTANCE was a particularly strong indicator of this factor, whereas ONLINESHOPFREQ demonstrated modest loading, indicating greater variability in the frequency with which consumers engage in online shopping. This observation aligns with recent research highlighting the wide variation in digital shopping habits, even among consumers who prioritize the convenience and extensive availability of online options (Verhagen & van Dolen, 2009). However, in the structural model, L3 exhibited only weak or non-significant effects on REGRETFREQ when other variables were controlled. This finding contrasts with previous studies that have directly linked online shopping to increased impulsivity and post-purchase regret, attributed to ease of access, persuasive interfaces, and limited tactile evaluation (Dholakia, 2000; Verhagen & van Dolen, 2009). One interpretation is that the psychological and behavioral mechanisms that lead to regret in online contexts, such as inadequate information search or susceptibility to promotions, are more accurately captured by other constructs in the model (e.g., splurging, deal proneness, or general risk-taking) than by a general orientation toward online availability. In essence, being an "online-ready" consumer does not inherently result in more frequent regret; rather, regret appears to be contingent on how online tools are utilized within the broader decision-making framework.

Collectively, these findings largely corroborate and refine the extant literature on consumer regret. Initially, they affirmed that symbolic, status-oriented motives (L1) are inclined to increase the likelihood of regret, whereas robust decisional and financial agency (L2) consistently mitigate it, aligning with theories that highlight the influence of self-attribution, counterfactual thinking, and confidence in shaping regret responses (Inman & Zeelenberg, 2002; Zeelenberg & Pieters, 2007). Furthermore, they indicate that technological transitions toward online commerce (L3) do not inherently lead to more frequent regret when psychological dispositions are considered, thereby complicating simplistic narratives about digital markets as inherently inducing regret. From a consumer behavior perspective, the findings suggest that interventions aimed at reducing buyers' remorse should focus less on discouraging online shopping and more on fostering realistic expectations, enhancing decisions and financial literacy, and distinguishing intrinsic from status-driven purchase motives. Assisting consumers in aligning their purchases with stable, quality- and value-based criteria, rather than with fluctuating social comparisons, may be particularly effective in diminishing the incidence of consumer regret.

## 6. Implications for Marketers and Management

Studies in this area provide crucial insights for marketers and managers who aim to reduce the negative effects of consumer regret. The primary strategy is to lessen the elements that cause regret: the perceived value gap between chosen and unchosen options and the consumer's feeling of responsibility (Tsiros & Mittal, 2000). For marketers, this implies a focus on post-purchase reassurance to ease cognitive dissonance and regret. Tactics include sending follow-up communications that emphasize the positive features of the purchased item, offering social proof (e.g., "Other customers who bought this also loved it"), and providing outstanding customer service to lower the perceived risk of the decision (Özyörük, 2022). In the digital space, supplying comprehensive and accurate product details and high-quality images helps align pre-purchase expectations with post-purchase reality, thereby reducing the likelihood of regret (Li, 2021). Management should focus on developing effective policies. A flexible return policy can be a double-edged sword; while it reduces anticipated regret and encourages initial purchases, it must be balanced against the operational costs of high return rates (Liu, 2025). Management can also use regret-priming marketing strategies that highlight the potential regret of not purchasing, thus shifting the focus from post-purchase regret to pre-purchase opportunity regret (Nasiry, 2012). Ultimately, addressing consumer regret is a strategic imperative that requires a holistic approach that integrates marketing communication, product information quality, and customer-centric return policies to foster long-term satisfaction and loyalty.

## 7. Directions for future research

Future investigations into how often consumers experience regret should delve deeper into the evolving and situational aspects of regret during the decision-making process. This involves moving past cross-sectional self-reports and adopting longitudinal and experience-sampling methods that capture regret episodes in real-time (Saffrey, Summerville, & Roese, 2008). Such research could explore the interaction between temporary conditions (such as mood, scarcity signals, or social comparisons) and more stable characteristics, such as decision making/financial autonomy and status orientation, which influence both the emergence and intensity of regret (Inman & Zeelenberg, 2002; Zeelenberg & Pieters, 2007). Another important area of focus is examining how digital settings, such as algorithmic suggestions, review sites, dynamic pricing, and social media, either heighten or lessen regret by increasing counterfactual comparisons and exposure to post-purchase information (Huang, Korfiatis, & Chang, 2018; Tsiros & Mittal, 2000). Additionally, cross-cultural and developmental studies are necessary to comprehend how factors such as age, socioeconomic background, and cultural norms related to responsibility and consumption influence chronic regret patterns throughout a person's life (Gilovich & Medvec, 1995). Finally, future research should more explicitly connect the frequency of regret to subsequent behavioral outcomes, such as complaints, switching, word-of-mouth, and long-term brand avoidance, by using comprehensive structural models that incorporate both individual processes (e.g., rumination, coping) and marketplace reactions (Chebat, Davidow, & Codjovi, 2005; Zeelenberg & Pieters, 2004). Together, these research directions would shift the literature from static descriptions to a more process-oriented and context-aware understanding of when and why regret becomes a persistent aspect of the consumer experience.

## 8. Limitations

This research identifies several limitations that should be taken into account when interpreting the findings and their implications for consumer regret. Firstly, although online panels and crowdsourcing platforms offer a cost-effective and rapid means to access diverse samples, they are non-probability samples and may not accurately reflect the broader consumer population (Goodman, Cryder, & Cheema, 2013). Participants on these platforms tend to be more digitally adept, more accustomed to online tasks, and potentially more focused on prices and deals than the average consumer, which could distort both the levels of regret and the patterns of relationships among status signaling, decision agency, and readiness for online commerce. This reliance on self-selected online participants restricts the external validity and generalizability of the results to offline or less digitally engaged groups. Secondly, the study utilized cross-sectional self-report data, limiting its capacity to infer causality. Regret (REGRETFREQ), status signaling orientation (L1), decision-making confidence, financial prudence (L2), and online commerce readiness (L3) were all measured at a single point in time, hindering the determination of temporal precedence. While the structural equation model supports a directional interpretation, such as the notion that greater decision/financial agency reduces subsequent regret, it is equally plausible that ongoing regret affects perceived decision competence or financial self-efficacy over time (Zeelenberg & Pieters, 2007). Longitudinal panel designs or experimental manipulations of choice environments would provide stronger evidence for causal pathways (Kline, 2016). Thirdly, a related limitation involves common method variance and biases from self-reporting. All key constructs were collected through self-reports during a single survey session, which might have inflated correlations due to shared method variance, social desirability, or the respondent's mood at the time (Podsakoff et al., 2003). For example, individuals with a generally negative mood might report higher levels of regret and lower confidence in their decisions, regardless of their actual actions. Future research could incorporate behavioral indicators of regret (e.g., return behavior, complaint rates), objective financial outcomes, or reports from others to cross-verify self-reported regret and decision-making agency. Fourthly, limitations in the measurement strategy must be acknowledged. The model treated Likert-type indicators (e.g., status, optimism, decision confidence) as approximately continuous and used continuous SEM with robust corrections. Although this is a common practice, especially with 5-point scales, it may not fully account for the ordinal nature of the data (Rhemtulla, Brosseau-Liard, & Savalei, 2012). Similarly, the decision to specify relatively simple reflective measurement models for L1–L3 might overlook potential multidimensionality; for instance, status signaling could reasonably be divided into social display versus brand-centric subdimensions. Fifth, from a modeling standpoint, the research employed a single global SEM where all predictors were included at once. While this approach is statistically efficient, it may mask important variations. The study did not explore nonlinear effects, interactions (such as those between status orientation and decision agency), or latent consumer groups with unique regret profiles. Other methods, like latent class SEM, finite mixture modeling, or multigroup SEM (e.g., segmented by age or income), could help identify subgroups where status, online behavior, and decision confidence operate differently (Kline, 2016; Nylund-Gibson & Choi, 2018). The reliance on REGRETFREQ as a sole global item is a notable limitation. Although this measure fits the theoretical concept of regret as a recurring emotional reaction to past purchase choices (Inman & Zeelenberg, 2002; Zeelenberg & Pieters, 2007), single-item measures are less reliable and do not capture the complex nature of regret, which includes dimensions like outcome regret, process regret, self-blame, and opportunity-based regret. Using multi-item regret scales or scenario-based vignettes could provide a more detailed measurement and allow for separate modeling of different regret aspects and their causes (Tsiros & Mittal, 2000). Sixth, contextual elements such as macroeconomic uncertainty, inflation awareness, and recent financial shocks may significantly impact regret experiences. In conclusion, future research using probability-based or more varied samples, longitudinal or experimental designs, richer multi-item regret measures, and alternative modeling techniques for ordinal and diverse data would be advantageous in confirming and expanding the current findings.

## 9. Funding

This research did not receive any specific grant from funding agencies in the public, commercial, or not-for-profit sectors.

## 10. Conflict of interest

The authors have declared no conflict of interest.